%% file: 00_paper.tex
\documentclass[acmsmall,screen,nonacm]{acmart}

\usepackage[shortlabels]{enumitem}
\usepackage{amsmath}
\usepackage{xcolor}

\newcommand{\edit}[1]{{#1}}

\AtBeginDocument{%
  \providecommand\BibTeX{{%
    \normalfont B\kern-0.5em{\scshape i\kern-0.25em b}\kern-0.8em\TeX}}}

\setcopyright{acmlicensed} 
\copyrightyear{2022}
\acmYear{2022}
\acmDOI{10.1145/1122445.1122456}

\acmConference[CHI Play '22]{Bremen, Germany '22: CHI Play}{November 02--November 05, 2022}{Bremen, Germany}
\acmBooktitle{Bremen, Germany '22: CHI Play,
  November 02--November 05, 2022, Bremen, Germany}
\acmPrice{15.00}
\acmISBN{978-1-4503-XXXX-X/18/06}

\acmSubmissionID{9248}

\begin{document}

\title{ggViz: Accelerating Large-Scale Esports Game Analysis}

\author{Peter Xenopoulos}
\email{xenopoulos@nyu.edu}
\affiliation{%
  \institution{New York University}
  \city{New York}
  \state{NY}
  \country{USA}
}

\author{João Rulff}
\email{jlrulff@nyu.edu}
\affiliation{%
  \institution{New York University}
  \city{New York}
  \state{NY}
  \country{USA}
}

\author{Claudio Silva}
\email{csilva@nyu.edu}
\affiliation{%
  \institution{New York University}
  \city{New York}
  \state{NY}
  \country{USA}
}


\begin{abstract}
\input{01_abstract}
\end{abstract}



\begin{teaserfigure}
  \includegraphics[width=\textwidth]{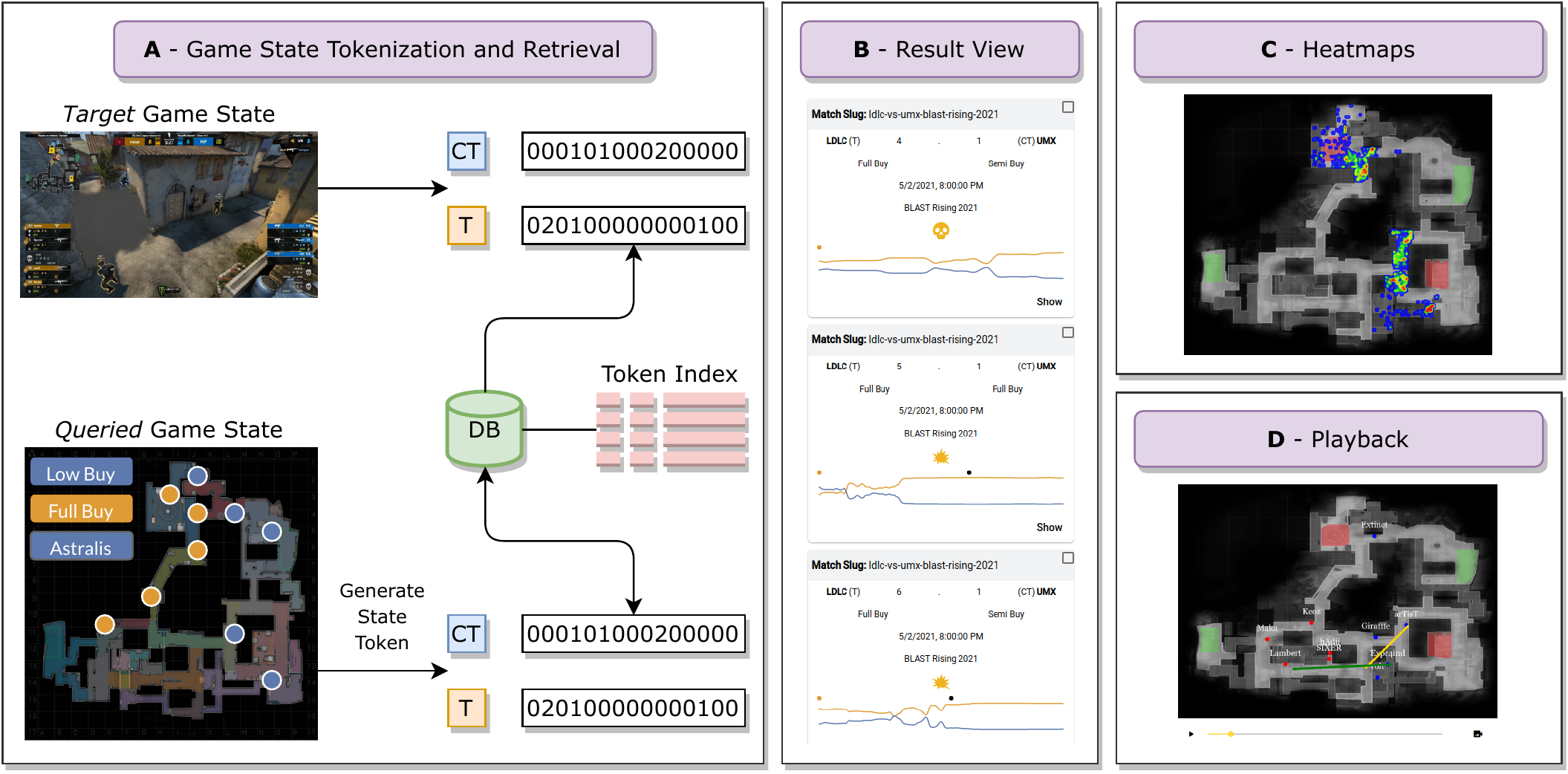}
  \Description[ggViz system teaser]{The ggViz system consists of four major components, a game state tokenization and retrieval framework, the result view showing the set of similar game states, a heatmap visualization to summarize the set of similar game states and a playback view to view Counter-Strike rounds in 2D.}
  \caption{We introduce a performant tokenization methodology to quickly summarize esports game states \edit{(A)}. Game states are assigned tokens according to the spatial distribution of players and a discretization of the world. Users can query game states by sketching player positions in the \textit{ggViz} interface, which queries a database created from a collection of thousands of professional esports matches. Users can view the most similar game states to their query \edit{(B)} and observe the win probability for the round in which the game state occurred. A user can either generate heatmaps of common player locations in the query's result set \edit{(C)} or watch detailed replays of specific rounds \edit{(D)}. We achieve interactive performance across 10+ million game states and evaluate ggViz with top esports teams.}
  \label{fig:teaser}
\end{teaserfigure}

\maketitle

\input{02_introduction}
\input{03_related_work}
\input{04_requirements}
\input{05_play_retrieval}
\input{06_system}
\input{07_evaluation}
\input{08_discussion}
\input{09_conclusion}


\begin{acks}
This work was partially supported by NSF awards CCF-1533564, CNS-1544753, CNS-1730396, CNS-1828576, and CNS-1229185. P. Xenopoulos and C. Silva were also funded by Capital One. C. Silva is partially supported by DARPA. Any opinions, findings, and conclusions or recommendations expressed in this material are those of the authors and do not necessarily reflect the views of DARPA.
\end{acks}

\bibliographystyle{ACM-Reference-Format}
\bibliography{00_bib}


\end{document}

%% file: 01_abstract.tex
While esports organizations are increasingly adopting practices of conventional sports teams, such as dedicated analysts and data-driven decision making, video-based game review is still the primary mode of game analysis. In conventional sports, advances in data collection have introduced systems that allow for sketch-based querying of game situations. However, due to data limitations, as well as differences in the sport itself, esports has seen a dearth of such systems. In this paper, we leverage player tracking data for Counter-Strike: Global Offensive (CSGO) to develop \textit{ggViz}, a visual analytics system that allows users to query a large esports data set through game state sketches to find similar game states. Users are guided to game states of interest using win probability charts and round icons, and can summarize collections of states through heatmaps. We motivate our design through interviews with esports experts to especially address the issue of game review. We demonstrate ggViz's utility through detailed cases studies and expert interviews with coaches, managers and analysts from professional esports teams.


%% file: 02_introduction.tex
\section{Introduction} \label{sec:introduction}

Although team-based esports have grown drastically over the past few years, there has been limited analytical adoption by esports teams and players~\cite{keiper2017no}. This limited analytical adoption is not for a lack of want -- video game players are increasingly seeking out visualizations that can aid in retrospective game analysis~\cite{DBLP:journals/tvcg/BowmanEJ12, DBLP:conf/chi/WallnerWBK21}. The lack of analytical tools is in part due to a lack of accessible data, which has typically fostered analytics in other sports~\cite{basole2016sports}. \edit{Accordingly, watching game replays is a common way to analyze matches}, whereby players, coaches and analysts systematically review both their opposition and their own recorded games to discover strengths, weaknesses and general patterns. 

\edit{Post-match review} can be an extremely time-intensive and manual process, especially for a team with limited resources, as a single game can stretch for many hours. Thus, to find game scenarios \edit{(states)} of interest, one must watch game \edit{replays} from start to finish. For media organizations, which may have resources to parse through thousands of videos to discover relevant highlights for broadcast, large databases can hold slices of videos tagged by keywords, although this approach may often lead to imprecise results~\cite{sha2016chalkboarding}. \edit{Although many tools exist to replay and view game recordings, few, if any, contain game state retrieval functionality.} Ultimately, for many stakeholders in esports, the ability to query for specific game situations would greatly improve productivity on an otherwise arduous and time-consuming workload.

Given manual game review's difficulty to scale, sports like baseball, soccer and basketball, have developed multiple data acquisition methods to allow for data-driven analysis. For example, there exist services \edit{which employ people to manually identify and record} events in sports games~\cite{liu2013inter}. Recently, computer vision-based systems, which automatically generate spatial and event data from video, are gaining traction~\cite{d2010review}. With large, clean and reliable spatiotemporal and event data, contemporary sports have developed visual analytics systems to allow analysts to query for game situations of interest, thereby greatly improving productivity in the game review process. These systems utilize the growing amount of player tracking and event data, which can be used to visualize portions of a game in a 2D interface in place of a video.

Since esports are played entirely virtually, one might expect data capture to be straightforward. However, the tools for data acquisition are surprisingly limited, and if they exist, are undocumented~\cite{charleer2018towards}. Furthermore, esports data is often stored in \edit{game logs} containing inconsistencies and lacking a commonly accepted data model~\cite{bednarek2017data}. Thus, most data in esports exists only as simple count-based statistics, as opposed to complex spatiotemporal and event data seen in other sports~\cite{maymin2018open}. These aforementioned data challenges have hampered development of game state querying and retrieval systems for esports, since player tracking and event data is difficult to acquire. 

This paper presents ggViz (Figure~\ref{fig:teaser}), a visual analytics system to navigate a large corpus of spatiotemporal Counter-Strike: Global Offensive (CSGO) data. We focus on CSGO as it is one of the most popular esports and has robust and \edit{public data} compared to other video games. ggViz allows users to search for game situations of interest through a sketch-based query interface. Powering ggViz is a fast and performant game state retrieval approach that efficiently encodes game scenarios and can return results on a data set of over 10+ million game states in under a few seconds. To do so, we exploit a game's navigation mesh, which is a graph of discrete spatial information that computer-controlled players (bots) use to move in-game. We generate tokens for each game state, and these tokens describe the approximate spatial distribution of players in the given game state. Our approach is game-agnostic, and can easily be extended to other esports or even conventional sports where the playing surface can be discretized, such as a soccer field or basketball court. We motivate ggViz's development through interviews with analysts, coaches and managers from top esports teams. We then evaluate ggViz through expert interviews and case studies. We show that ggViz can facilitate quick game state retrieval and is well received by professional esports experts. 
\\[1em]
\textbf{Contributions.} We make the following contributions:

\begin{enumerate}
    \item A game-agnostic game state retrieval framework which summarizes the spatial distribution of players in a given game state through a fast tokenization algorithm.
    \item ggViz, a visual analytics system that allows users to sketch game states of interest to query for similar game states.
    \item An evaluation of ggViz through expert interviews, including coaches, analysts, and managers from top esports teams, and expert-inspired use cases.
\end{enumerate}

The rest of the paper is structured as follows. Section~\ref{sec:related-work} reviews relevant literature on game states, esports visualization and conventional sports play retrieval. In Section~\ref{sec:domain-req} we discuss Counter-Strike, its data, and conduct a requirements analysis inspired by interviews with esports analysts, coaches and managers. Section~\ref{sec:play-clustering} outlines out game state retrieval methodology. In Section~\ref{sec:system}, we present the ggViz system and describe each of its components. Section~\ref{sec:evaluation} contains the results of an expert study, as well as two use cases motivated by expert workflows. Finally, we provide a discussion of our work in Section~\ref{sec:discussion} and conclude the paper in Section~\ref{sec:conclusion}.

%% file: 03_related_work.tex
\section{Related Work} \label{sec:related-work}

\edit{
\subsection{Gameplay Visualization} \label{sec:gameplay-vis}
Gameplay visualization is a long standing field and is particularly relevant to game designers and players. Gameplay data visualization often necessitates visualizing many features at once. Another important component of gameplay data are the spatial features, since many games have a fundamental spatial aspect to them~\cite{DBLP:conf/cig/DrachenS13}. Oftentimes, games are analyzed and visualized through their states, such as with Playtracer~\cite{DBLP:conf/fdg/AndersenLABP10}. However, the main weakness of Playtracer is that it is unusable for games with continuous state spaces, like many esports. Liu~et~al. address this weakness by considering game state features~\cite{DBLP:conf/fdg/LiuASCP11}. 

Wallner and Kriglstein develop a system to visualize a game space as nodes that a player visits over time~\cite{DBLP:conf/chi/WallnerK12}. Each node represents a game state that is defined by a feature vector and they cluster game states and visualize information within each cluster. They perform a case study using Team Fortress 2 and provide findings on weapon usage, kill locations and map control. Next, Wallner proposes to represent gameplay through a graph to analyze and visualize gameplay data\cite{DBLP:conf/fdg/Wallner13}. Finally, they build upon the aforementioned works by proposing PLATO, a visual analytics system which takes advantage of graph analysis techniques to analyze gameplay data~\cite{DBLP:journals/cg/WallnerK14}. Ahmad~et~al.~present \textit{Interactive Behavior Analytics} (IBA), a methodology to model player behavior in Dota 2~\cite{AhmadBKTNE19}. Specifically, they introduce StratMapper, a visualization system to playback game events. StratMapper also allows for user labeling of higher-order events. Kleinman~et~al. extend the aforementioned work by incorporating sequence analysis in conjunction with StratMapper~\cite{DBLP:conf/fdg/KleinmanATBNHE20}.

Summary gameplay visualizations have proven useful in other games in the form of maps for post hoc game analysis. Inspired by military battle maps, Wallner and Kriglstein develop \textit{battle maps} for team-based combat games~\cite{DBLP:conf/chiplay/WallnerK16}. Wallner extends the aforementioned work into an automatic battle map generator for World of Tanks~\cite{Wallner18}. Similarly, for StarCraft II, Kuan~et~al. develop a system that allows for user to analyze engagements from both global and local views~\cite{DBLP:conf/ieeevast/KuanWC17}. Wallner and Kriglstein propose an approach to visualize multivariate game information using hexbin maps~\cite{DBLP:conf/chiplay/WallnerK20a}. They apply their approach to data from StarCraft: Brood War and find they could display one to four features in the glyph contained in each bin.

Gameplay visualization can also be useful to understand player behavior. For example, Li~et~al.~develop a visual analytics system to identify specific game occurrences in a massive online battle arena game (MOBA)~\cite{LiXCWWQM17}. MOBAs, like Defense of the Ancients 2 (Dota 2) or League of Legends (LoL) are popular esports. Li~et~al. extend this work by incorporating machine learning based models to recommend interesting portions of a match~\cite{LiWXQM18}.
}

\subsection{Esports Visualization}
\edit{
Esports visualization is a growing field, particularly concerning applications geared towards strategy identification and understanding \edit{player tendencies}~\cite{KriglsteinMVKRT21}. Although many of the approaches and systems in Section~\ref{sec:gameplay-vis} are not designed nor evaluated with esports in mind, they may easily be extended or used for game summarization or playback in esports. Much existing esports-specific visualization work considers applications geared towards media production and spectating. Increasingly, we are also seeing more work targeted towards esports match analysis, particularly for popular MOBA and FPS esports, like Dota 2, LoL and CSGO. However, there has been limited work towards visual interfaces for game state retrieval within esports.

Concerning media production, Block~et~al.~introduce \textit{Echo}, a production tool used to detect extraordinary player performances to translate into audience-facing graphics~\cite{BlockHHSDUDC18}. Echo was developed specifically for Defense of the Ancients (Dota 2) and focused on displaying key performance indicators for players. Charleer~et~al.~designed informational dashboards for a variety of games including LoL and CSGO~\cite{CharleerGGCLV18}. They found that dashboards were helpful for spectators, but there were tradeoffs and considerations when it came to the cognitive load of those spectators. Finally, Kokkinakis~et~al.~propose \textit{Weavr}, a companion app that allows audiences to consume data-driven insights during a match. They found that users had a strong interest in consuming analytical esports content~\cite{KokkinakisDNOPR20}.

For game analysis, esports visualization efforts have been directed towards game summarization or towards understanding spatial characteristics of a game through game playback. Indeed, Wallner~et~al. show that spatial data is sought-after information by players in post-game visualizations~\cite{DBLP:conf/chi/WallnerWBK21}. Many esports visualization works focus on MOBAs. For example, Gonçalves~et~al. and Afonso~et~al.~present VisuaLeague I and VisuaLeague II, respectively~\cite{GoncalvesV0CM18, Afonso19}. These systems, designed for LoL analysis, provide match playback capabilities. They found that users were receptive to animated 2D replays. Recently, Weixelbaum and Matkovic introduce Rumble Flow++, a visual analytics system used to explore Dota 2 games by visualizing interactions between players using graphs~\cite{weixelbaum2021rumble}. 

While most of the aforementioned esports visual analytics works have focused on MOBA games, there is a dearth of FPS-oriented esports visualization work. Horst~et~al.~introduce CS: Show, an FPS match analysis tool designed for CSGO~\cite{DBLP:conf/iwec/HorstZD21}. In their tool, users can play back and summarize rounds from a single CSGO match. However, while CS:Show, it is not equipped to analyze a corpus of CSGO match data, nor does it contain retrieval functionality. 

We build upon prior work in the following directions. Many of the prior gameplay and esports visualization systems focus on game summarization or playback. However, there has been relatively little work directed towards game state retrieval, particularly for esports, as most of the existing game state retrieval is built with board games in mind~\cite{DBLP:conf/sigir/GangulyLJ14, DBLP:conf/sigir/UshikuMKT17}. We address the issue of esports game state retrieval in close cooperation with esports experts from design to evaluation. Additionally, while many of the aforementioned works are built for use with a single match. In our work, we focus on analyzing a corpus of matches.
}

\subsection{Sports Play Retrieval}
As many esports teams continue to closely resemble conventional sports teams, we can use lessons learned from the latter. Play and state retrieval, especially through sketching, is \edit{well-}established in conventional sports, as much of the sports workflow for analysts involves reviewing \edit{game replays} to find scenarios of interest~\cite{clegg2020data}. \edit{Sketch-based querying has gained traction in other fields}, such as time series search~\cite{siddiqui2020shapesearch, mannino2018expressive}, image retrieval~\cite{zhang2016sketch} and user interface discovery~\cite{huang2019swire}.

Shao~et~al.~first introduced a system for searching trajectory data in soccer by means of an interactive search interface that allows a user to sketch trajectories of interest~\cite{shao2016visual}. \edit{Their play retrieval procedure creates a feature vector by discretizing portions of the field and assigning a value where the trajectory crosses. Users can draw trajectories of interest. Stein~et~al.~expand upon the previous work by considering more context around the play and further improve the sketch-based visual system~\cite{stein2019tackling}. Sha~et~al. introduced \textit{Chalkboarding}, a query paradigm and retrieval system which allows users to draw plays of interest in basketball~\cite{sha2016chalkboarding}. Each play is also assigned a cluster based on ball movement during the play. To facilitate a quick search speed, the queried play is only considered against those in the same cluster. The authors conducted a user study which showed great improvements over a keyword-based play search system. Sha~et~al.~(2018)~showed how an interface for Chalkboarding can be used for broadcast purposes~\cite{sha2018interactive}. Di~et~al.~extended upon Chalkboarding by learning a user-specific model that ranks play search results based on search result click behavior~\cite{di2018large}.}


While play retrieval has received increasing interest in sports analytics, to the best of our knowledge, there is no current work specific to esports play retrieval. To a large degree, esports differ from the sports in the aforementioned works, which revolve mostly around basketball and soccer. Esports have a much more decentralized structure than conventional ball-based sports. For example, esports lack a ball to serve as a focal point, nor do players have well-defined positions. Accordingly, applying methods like Chalkboarding, which rely on clustering ball movement, are hard to apply. Finally, since in many esports, such as CSGO, players stay stationary for long periods of time, we may be interested in \textit{states} (positions) rather than \textit{trajectories} (movement). 

%% file: 04_requirements.tex
\section{Data and Domain Requirements} \label{sec:domain-req}
\edit{In our study, we focus on Counter-Strike: Global Offensive (CSGO), a popular esport. We focus on CSGO for two reasons. The first is that CSGO has an extensive player base. At the time of writing, CSGO had rougly one million peak players. Furthermore, CSGO has an extensive esports infrastructure, with many tournaments and leagues, as well as teams with salaried players. Secondly, CSGO data is one of the few esports in which one can easily obtain detailed spatiotemporal game data. This game data is available through game replay files, which are available publicly on fan sites. We describe CSGO in Section~\ref{sec:csgo-description} and its data in Section~\ref{sec:csgo-data}. In Section~\ref{sec:domain-req}, we conduct interviews with CSGO analysts, coaches and managers to derive a list of requirements that our system should address.}

\subsection{Counter-Strike Description} \label{sec:csgo-description}
Counter-Strike: Global Offensive is a first-person shooter (FPS) style video game where two teams of five players compete across a variety of objectives. A professional game takes place over one or more \textit{maps}, which are distinct virtual worlds. Most professional CSGO games are structured as a best of three maps. There is a pool of seven maps for competitive play, so to determine the three maps played in a game, each team participates in a picking and banning process before the game to determine which maps are played. On each map, teams are initially assigned a side, either the Terrorists (T) or the Counter-Terrorists (CT), and then play for 15 rounds as their assigned side. The two teams switch sides after 15 rounds. A team wins a map by winning 16 rounds.

Both the T and CT sides can win a round by completing a variety of objectives, such as eliminating all members of the opposing side, having the bomb explode or be defused, or running out of time. Both sides can win a round if they eliminate all members of the opposing side. The T side can win a round by planting and exploding a bomb at one of the two bombsites on a map, denoted A or B. At the start of each round, one T player is assigned the bomb. Once planted, the bomb explodes in 35 seconds, unless defused by the CT side. If the CT side successfully defuses the bomb, they win the round. If no win condition is met by the end of the round time (which is roughly two minutes), the CT side wins by default. Players start each round with 100 health points (HP) and are eliminated from a round when they reach 0 HP. Specifically, players lose HP when they are damaged -- typically from gunfire and grenades from the opposing side. Players buy equipment, such as guns, grenades (also called utility) and armor, at the beginning of a round, using virtual money earned from doing well in previous rounds. 

\subsection{CSGO Data} \label{sec:csgo-data}

\begin{figure}
    \centering
    \includegraphics[scale=0.7]{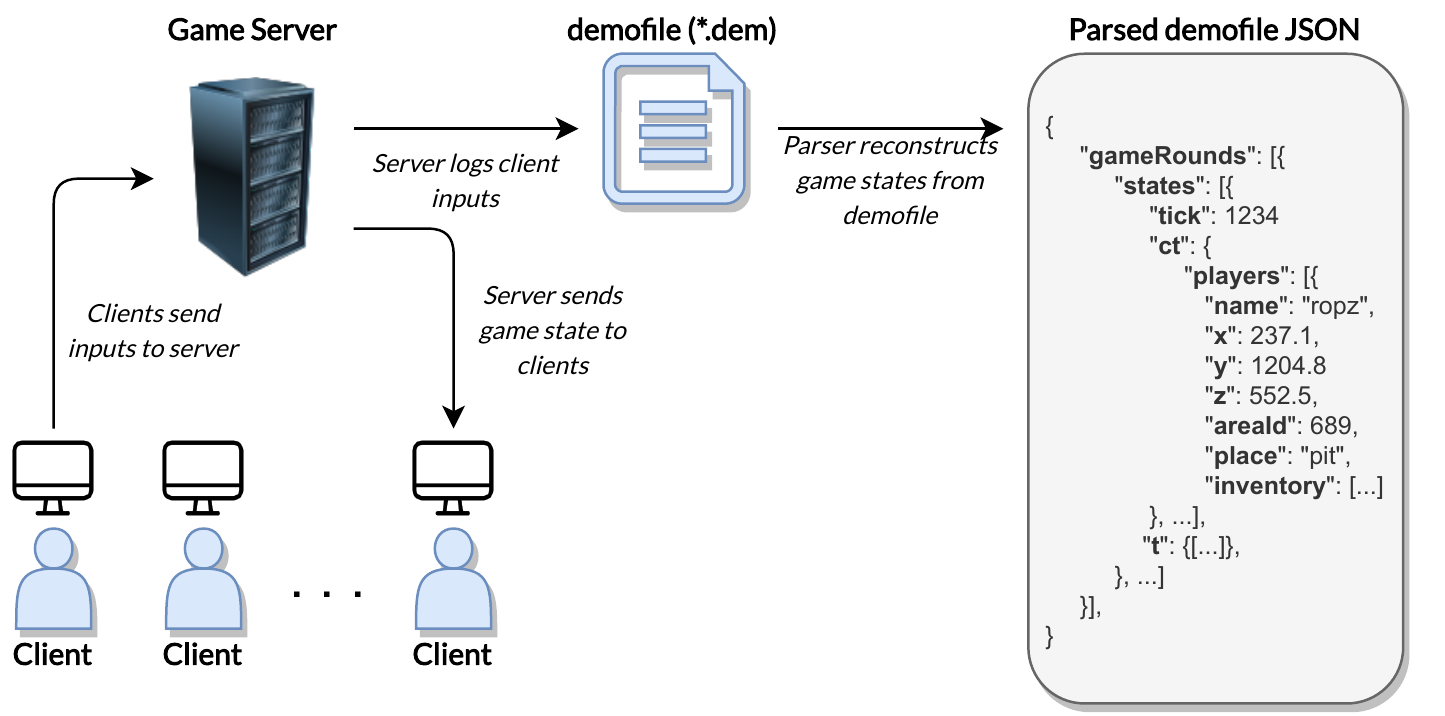}
    \Description[CSGO data generation]{Esports are often structured in a client-server environment. We parse CSGO demo files, which record the game server updates.}
    \caption{Clients (players) provide input through keystrokes and mouse clicks, which changes their local game state. Clients then send these local game states to the game server, which reconciles the client states and sends a global game state back to all clients. While doing so, the server writes the global state to a demo file, which we then parse to generate a JSON containing spatiotemporal information about a match, including player events and trajectories.}
    \label{fig:demofile_generation}
\end{figure}

CSGO games take place in a client-server architecture. The clients (players) connect to a game server where they play their games. Client inputs, such as keyboard presses and mouse clicks, are recorded by the server and resolved across all clients every \textit{tick}. In competitive play, there are 128 ticks in a second, meaning a tick is roughly 8 milliseconds. Additionally, server side changes, such as round starts and ends, or bomb explosions, are also resolved across all clients. We show how these events are generated, and their parsed output, in Figure~\ref{fig:demofile_generation}.

A game can be recorded by the server through the use of a \textit{demo file} (\texttt{.dem}). The demo file contains a serialization of all client-server communications. When a demo file is recording, it writes data in tick bursts, which contain around eight ticks~\cite{bednarek2017data}. While a demo file can be recorded by a client, we focus solely on server-recorded demo files, as they tend to be the predominant type of publicly available demo files. Server-recorded demo files are often released by tournaments and competition organizers on game community sites. Additionally, players can access a service called \textit{GOTV} in-game, which broadcasts real-time games through streaming demo files as they are written by the server.

Since the demo files are effectively unstructured log files, it is important to impose a data structure on them to better facilitate data analysis. To do so, we use a demo file parser that can translate demo files into an interpretable JSON format~\cite{XenopoulosWPACSGO}. The parser parses the game into a series of consecutive \textit{game states}. Let $G_{r}^t$ be a game state in round $r$ at time $t$. Each game state contains the complete set of information of the world at time $t$. Thus, a game state contains data such as player locations, health, armor and grenade positions. 

\subsection{Domain Requirements}
We conducted interviews with \edit{four} primary esports stakeholders, including coaches, analysts and managers from professional CSGO teams, to understand how they review past games. A1 and A2 are analysts, C1 is a coach and M1 is a manager. Each participant is employed by a \edit{separate} professional esports organization, and has at least three years of experience in esports (median of five years). At the time of writing, A1, A2 and M1 were part of top 10 ranked teams, and C1 was part of a top 50 ranked team.  

\edit{We asked each expert a series of semi-structured questions about how their CSGO demo analysis workflow is structured, what they tend to analyze in demos, what tools they use to analyze demos, and what difficulties they face in the demo review process, with particular emphasis on game scenario search. Each interview took place over Zoom. We analyzed the experts' responses and identified three main themes in the experts' responses. First, the experts outlined similar workflows to each other, and these workflows revolved around analyzing their opponent's performance in different game scenarios. Generally, this represented an attempt by the experts to understand their opponents strategically. Second, the experts' found great difficulty in finding appropriate tools to analyze matches. Third, a common difficulty was found in the ability to find scenarios of interest. As experts had to manually search demos to find scenarios of interest, demo analysis was typically a long and laborious activity.
}


The stakeholders unanimously expressed that they watch and review game replays regularly, specifically for match preparation. \edit{A shared approach to match preparation, identified by each expert, involved the expert along with other coaching staff manually reviewing recent demo files of their opponents on the maps which they are most likely to play through the use of a demo playback tool. The stated goals of viewing these demos was to understand how their opponents play.} Specifically, A1 said he looks for setups of a team under different parameters, such as how teams defend bombsites given specific equipment values or with different types of grenades. Similarly, C1 stated that their team's workflow revolved around summarizing setup tendencies of opponents under different circumstances. For example, answering a question like ``given that a team does not have much money, how do they defend a bombsite?'' is especially important. The \edit{common} workflow between experts involved outlining a few common game scenarios, typically stratified by equipment value, and then verifying how their opponent would defend or attack a bombsite given a specific game scenario. To do so, they would simply watch a round and make note of both the parameters and the outcomes.  

A1 and M1 used the CSGO in-game replay viewer to review demo files. Additionally, all interviewees also used industry-standard demo file viewers, but each agreed that there was still much work to make these viewers complete. \edit{For example, due to the sequential nature of CSGO demos, the in-game viewer can take many seconds to ``rewind'' to previous points of a demo, as the demo must be read from the beginning.} Third-party demo review tools \edit{directly address this issue and often have intuitive interfaces to navigate demo files and easily switch between rounds and game states.} Although these third-party tools are often not as visually similar as the actual game, they are much easier to use than the in-game viewer. In fact, A2 said that his team eschewed the in-game viewer in favor of third-party tools and plugins. \edit{Despite third-party tools, their lack of play retrieval functionality is still a problem and requires users to watch demo files manually to find scenarios they want.}  Because of this, each stakeholder remarked that reviewing demos can be an arduous and slow process, which prohibits any sort of large scale analysis. No stakeholder had a defined or algorithmic process to find similar plays or setups, and relied on either memory or manual search. 

Specifically, A1, like A2, mentioned that the hardest part of the scouting process was going through demo files and finding specific instances to clip or screenshot as demonstration to players and coaches. Furthermore, A1 mentioned that he seeks to find specific game scenarios ``all the time'', and that a tool to do so would be incredibly beneficial to his workload. Specifically, A2 said that sharing points of a match can be difficult when using demo file viewers, and that 2D reconstructions were sometimes easier to share. M1 said that the ``tools aren't there to find specific spatial setups'', and for that reason, demo analysis can become quite broad and be overloading for both analysts and players. C1 said that the hardest part of scouting and game review was ``finding certain situations that are relevant to you, as you do not have the time to look through ten demo files to find enough examples''. In particular C1 stressed his frustration with the ``anecdotal'' approach to game review that he felt was standard in the CSGO esports community.

From these responses, a system which can return accurate spatial setups in a fast manner could greatly improve these esports professionals' workloads. Furthermore, existing sports play retrieval systems revolved around retrieving \textit{trajectories}, and A1, C1 and M1 each added that they were instead seeking to search for particular defensive (CT) or offensive (T) positions rather than movement patterns. This is because teams often think about CSGO in terms of attacking and defending bombsites, and teams defend bombsites without much player movement.


From these interviews, we develop the following requirements:

\begin{enumerate}[start=1,label={\bfseries R\arabic*}]
\item \label{req:query_context} \textit{Query game scenarios by context.} In the CSGO game review, experts primarily watch matches with the goal of discovering \textit{strategies} that occur under specific \textit{conditions} \edit{(A1, A2, M1, C1)}. One of the strategies users are most interested in is understanding how teams defend bombsites \edit{(C1)}. In general, strategies are represented spatially through player locations \edit{(A1, M1, C1)}. Concerning game conditions, experts are especially concerned with finding opponent strategies in specific equipment level environments, such as when teams have good equipment or bad equipment \edit{(A1, C1)}. The proposed design must allow a user to query game states by player location and game conditions. Furthermore, most game review is done with a specific opponent in mind, meaning our system must also allow a user to query matches of a particular team. 

\item \label{req:result_metadata} \textit{Visualize game state metadata.} The game scenario search space is large, since a single CSGO game can be a few hours, and teams can play multiple times a week. One map has at minimum 16 rounds, and teams often play three maps a match, meaning matches can last multiple hours. Accordingly, users want to identify especially interesting scenarios without having to watch each scenario from the result set \edit{(A1, M1, C1)}. Thus, the proposed design should not only return results quickly, but also allow a user to easily identify interesting game scenarios returned in the result set. Wallner~et~al. stresses the importance of visualizing high-level knowledge in post-game scenarios~\cite{DBLP:conf/chi/WallnerWBK21}.

\item \label{req:playback} \textit{Playback for player trajectories and events.} Watching a demo file is important as doing so helps users gain the full context of a state in a given round. The proposed design should provide the ability for users to playback the round in 2D, since manually watching a game is a ubiquitous activity for our stakeholders \edit{(A1, A2, M1, C1)}. Additionally, player actions, like grenades and kills, should also be visualized in the playback, since they influence a situation's context. According to Wallner~et~al., spatial characteristics of the game were the most sought after data to visualize in post-game scenarios~\cite{DBLP:conf/chi/WallnerWBK21}.

\item \label{req:heatmap} \textit{Summarize query results.} Stakeholders are interested in summarizing tendencies of teams for presentation to other stakeholders, especially players. For example, C1 often displayed visuals and statistics to his players before games to identify and summarize the tendencies of the opposition. The system should be able to provide an intuitive visual summary that describes a team's strategy in a given context.
\end{enumerate}

%% file: 05_play_retrieval.tex
\section{Game State Retrieval} \label{sec:play-clustering}
As outlined by our requirement analysis in Section~\ref{sec:domain-req}, finding similar game states is a fundamental task in game review. For example, a coach may want to reference a particular strategy of an opponent, or a player may want to summarize their own positioning in a given context. In each case, it is important that queries not only return accurate results, but also do so in a timely manner. Accurate results are especially important in a field like sports analytics, where many users, like coaches and players, may be new to working with data or model outputs, and have little experience understanding faulty output. Speed is important as we seek to improve upon the slow game review and play retrieval process. Since player locations are a large component of a game scenario, a retrieval method which can efficiently represent player locations is crucial to powering visual analytics systems which query game scenarios.

Hashing, central to many existing play clustering methods, can be used to provide quick lookup for similar scenarios. If a state can be assigned to a cluster quickly, the search space can be markedly reduced. In other sports, plays and game states may be clustered by applying clustering algorithms, such as k-means (used in Chalkboarding~\cite{sha2016chalkboarding}), on ball movement. However, as CSGO, along with most other esports, is not a ball-based sport, such techniques, like those in Chalkboarding, cannot carry over. Furthermore, clusters can greatly be affected by the choice of parameters, \edit{such as the number of clusters in k-means, which may be hard to tune. Specifically, the returned cluster membership may not be tied to intrinsic game phenomena, but rather due to the choice of the $k$ parameter in k-means}. To solve the problem of quickly generating interpretable game state representations, we turn to computer-controlled \textit{bot} movement in video games.

\begin{figure}
    \centering
    \includegraphics[scale=0.6]{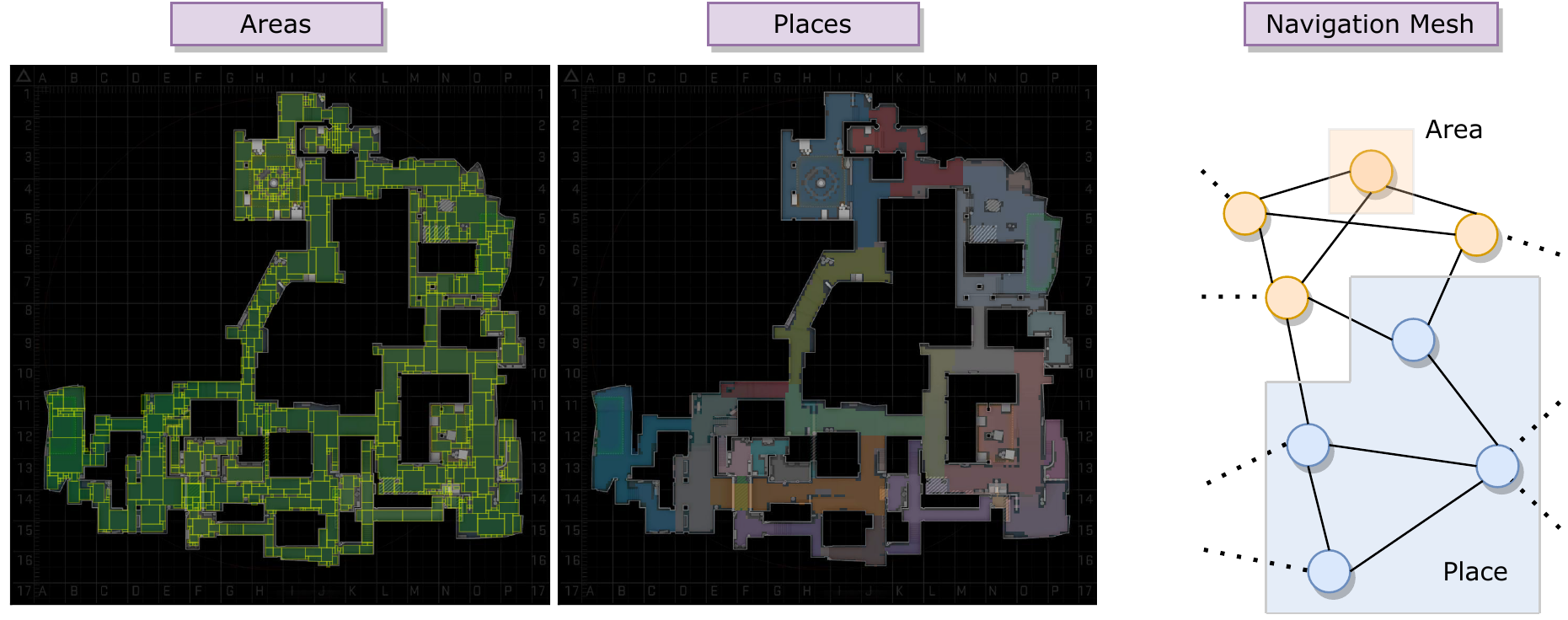} 
    \Description[Navigation meshes]{Navigation meshes are graphs that represent traversable surfaces in a 3D world.}
    \caption{Navigation meshes can be constructed as graphs. ``Areas'' are traversable surfaces (nodes), and a collection of areas forms a ``Place'' (set of nodes). Typically places are collections of nodes that are close to each other. Above, we visualize areas and places for popular CSGO map Inferno.}
    \label{fig:map_places}
\end{figure}

A bot is an computer-controlled player which can play against real players, and CSGO provides the functionality for players to play against bots. To move around the virtual world, bots use a \textit{navigation mesh}, $\mathcal{N}$, which provides information on all traversable surfaces of the virtual world~\cite{snook2000simplified, navigation}. Formally, a navigation mesh for map $m$ can be thought of as a graph where $\mathcal{N}_m = (A, C)$. $A$ is a collection of \textit{areas} (nodes), and $C$ is the set of connections between areas (edges). $A$ represents the set of traversable surfaces on a given map. Each area also belongs to a set of areas which is given a human-interpretable \textit{place} identifier. Each place has a name corresponding to a predefined region on the map, like ``Bombsite A'' or ``CT Spawn''. Thus, $\forall a_i \in A, \exists P_k \textrm{ such that } a_i \in P_k$, where $P_k$ denotes a human-interpretable place on the map. We show examples of areas and places extracted from navigation meshes for a few CSGO maps in Figure~\ref{fig:map_places}. The discretization of a playing space that a navigation mesh provides is not only found in video games, but also conventional sports as well. For example, some data providers discretize a playing field into zones to tabulate player fielding statistics. In soccer, it is common for coaches to discretize the pitch into different ``channels'' when talking about passing and formations. 

\begin{figure}
    \centering
    \includegraphics[scale=0.69]{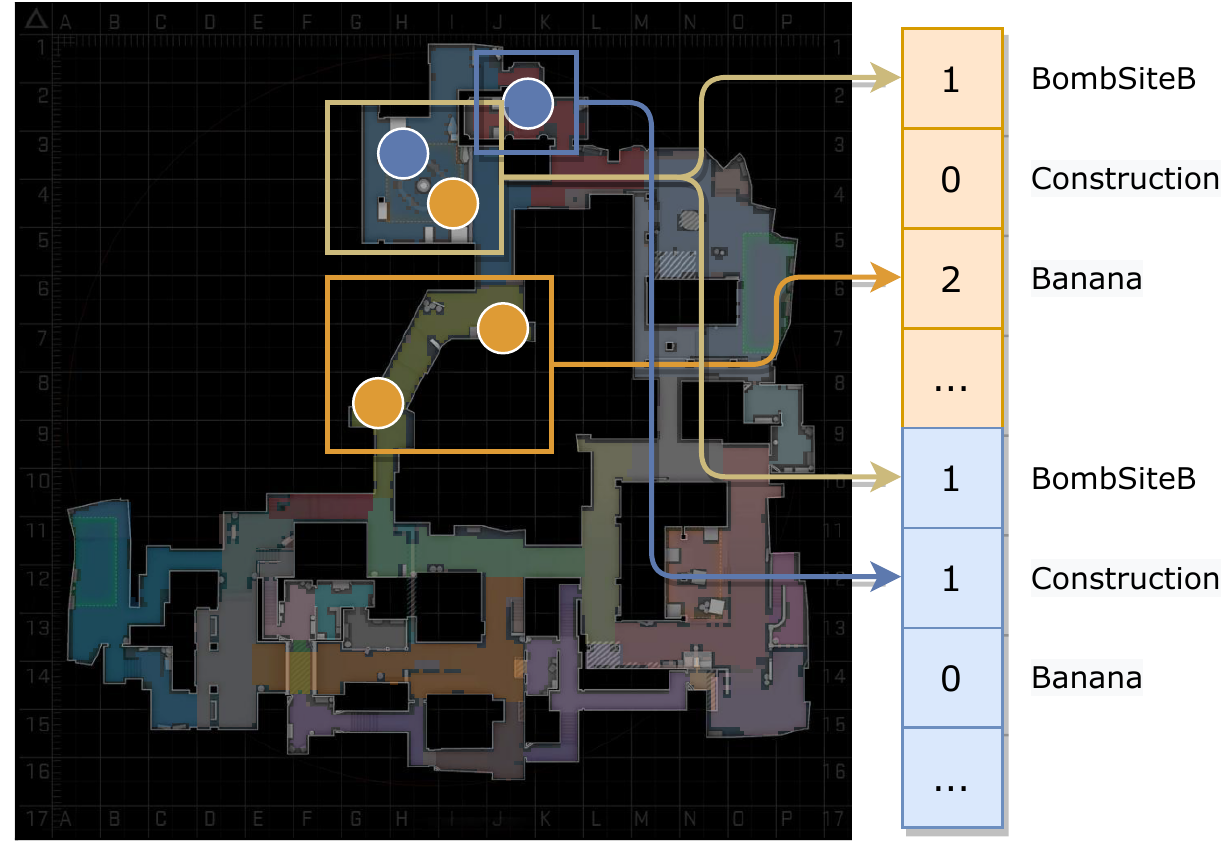}
    \Description[ggViz game state tokenization]{Our game state tokenization method counts the total number of players for each discrete navigation place for each side.}
    \caption{Our tokenization process for a game state on the \textit{Inferno} map. Each colored area on the map represents a distinct place. Orange represents the Terrorist side and blue represents the Counter-Terrorists. Place names are annotated under their position in the token string. Player counts are tabulated per side on each place and concatenate into a string.}
    \label{fig:play_tokenization}
\end{figure}

To facilitate quick retrieval, we draw from previous play clustering research, which utilizes hashing of similar plays, as well as from CSGO spatial data characteristics, particularly the navigation mesh described above. Our retrieval process starts with curating a large set of CSGO game states, which we call $\mathcal{S}$. Each game state $s$ includes player positions in the form of a set of $(x, y, z)$ coordinates. From the set of player coordinates, we can determine where each player is located in $\mathcal{N}$. For each state, we assign a token, using function $T(s)$. The generated token is simply a count of the number of players of each side in each place. The token is represented as a string of the form ``$P_1 \textrm{ } P_2 \textrm{ } P_3 \textrm{ } ...$'', where $P_k$ indicates the total player count in the nodes of $P_k$. We show an example of the token creation process in Figure~\ref{fig:play_tokenization}. We calculate a token for the T and CT sides. To create a single token for the overall state, we simply concatenate the two sides' tokens. Each token is indexed in a database for fast lookup. We can also save each individual place's player count, $P_k$, separately to allow for queries on portions of a token, such as queries where a user may want to find $X$ players in place $Y$.

Considering navigation mesh $\mathcal{N}$, game state set $\mathcal{S}$ and query (game state of interest) $q$, our retrieval method $R$ returns

\begin{equation}
    R(\mathcal{N}, \mathcal{S}, q) = \{s_1, s_2, ..., s_n\}
\end{equation}

\noindent where $T(q) = T(s_i), \forall i \in \{1, 2, ..., n\}$ and $T$ is our tokenization function. To query a large set of CSGO game states, a user can provide a list of player coordinates for both the Terrorists and Counter-Terrorists, which defines $q$. Providing these player coordinates can be accomplished via a visual interface where users can draw game states of interest, like in Chalkboarding~\cite{sha2016chalkboarding}. Then, we generate the token for $q$ by calculating $T(q)$. Using this token, our retrieval system queries a database containing the set of states $\mathcal{S}$. The number of candidate states can be further reduced by imposing non-spatial constraints, such as those on equipment value, number of grenades or specific teams. One can also construct other metrics to calculate intra-token similarity and rank returned game states. For example, one could find the distance to the closest player of the same side in $s_2$ for each player in $s_1$, given two game states $s_1$ and $s_2$. Then, the total distance between $s_1$ and $s_2$ can be the sum of these distances. This method clearly becomes intractable for queries which return large sets, or for searching over the entire database of game states, but it may be suitable for smaller sets of game states. To calculate inter-token similarity, for situations where the states where the generated query token may not exist, we can consider a modified Hamming distance, where the distance between two positions in a string is the absolute value difference between them. 

\edit{Our retrieval system consists of a preprocessing and retrieval step. In the preprocessing step, game states are assigned a token based as previously described. This token is stored as an indexed field in a database to facilitate quick lookup. To query similar game states, a sketch containing player coordinates are translated into a token representing player locations, and then the system searches the database for game states with the same token. Candidate game states are then ranked and displayed to the user.} Our approach has the following advantages: (1) representation of spatial player information is easily interpretable as it is directly tied to human-interpretable locations, (2) limited data to store beyond three tokens for each game state and (3) performance easily scales to millions of game states. Furthermore, our method is not beholden to using navigation meshes to discretize a playing surface. As our method only relies on \textit{some} discretization of a playing surface, our method is also sport-agnostic, and can easily be adopted by other esports and conventional sports. One important distinction with previous work on play retrieval is that prior systems focused mostly on trajectory similarity. As we see in Section~\ref{sec:system}, users in esports are often searching for specific \textit{scenarios} (game states), rather than trajectories (sequences of game states). 

%% file: 06_system.tex
\section{ggViz} \label{sec:system}

\subsection{System Design}

\begin{figure*}
    \centering
    \includegraphics[width=\textwidth]{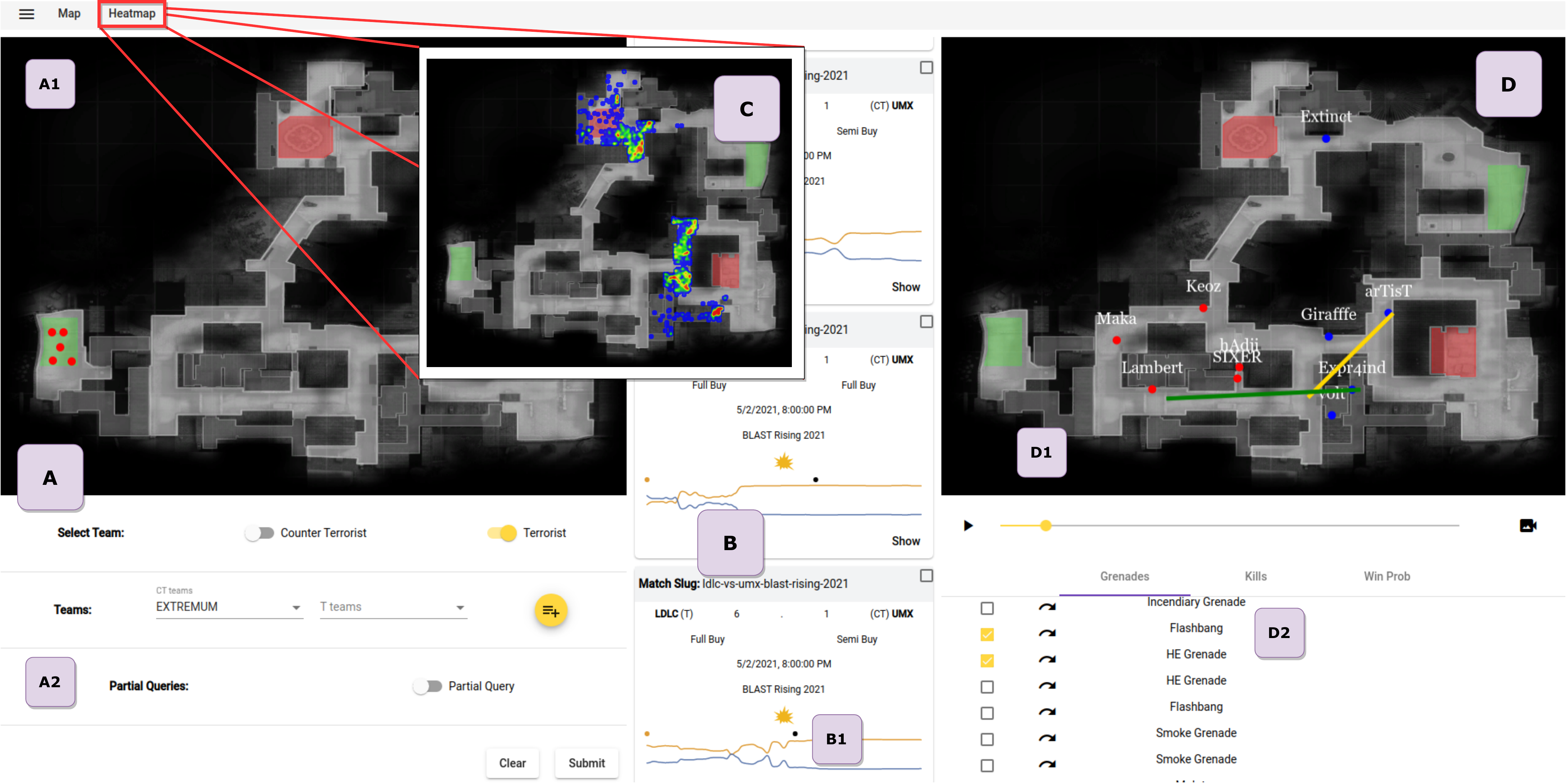}
    \Description[ggViz system]{ggViz has three main views, the query view, where users can draw game states of interest, the result view, where users view the set of similar game states, and the playback view, where users can watch the round in which a game state occurs.}
    \caption{ggViz system shown querying a game state. Users can query for game scenarios in the query view \edit{\boxed{\textbf{A}}} by sketching player positions \edit{\boxed{\textbf{A1}}} and applying team or round filters \edit{\boxed{\textbf{A2}}}. After querying, the result view \edit{\boxed{\textbf{B}}} shows a list of similar game scenarios. Users can navigate by looking at icons for how the round ended, or by observing win probability charts \edit{\boxed{\textbf{B1}}}. Users can also summarize the result set by viewing heatmaps of common CT/T positions for scenarios in the result set \edit{\boxed{\textbf{C}}}. Finally, users can play back the rounds that the returned game scenarios occur in through the playback view \edit{\boxed{\textbf{D}}}. The playback view shows both player trajectories \edit{\boxed{\textbf{D1}}} and important round metadata, such as grenade usage, kills and win probability \edit{\boxed{\textbf{D2}}}.}
    \label{fig:ggviz_system}
\end{figure*}

To fulfill the requirements identified by interviews with esports experts, we developed the ggViz system, coined after the popular gaming phrase ``good game (gg)''. In Figure~\ref{fig:ggviz_system}, we see ggViz and its constituent components. ggViz contains three components, (1) the \textit{query view}, to design a query across strategies and contexts, (2) the \textit{result view}, to observe the query result set and each element's associated metadata, and (3) the \textit{playback view}, to replay the round in which the game scenario occurred and identify key points of the round. \edit{All \boxed{\textbf{boxed}} references refer to Figure~\ref{fig:ggviz_system}.}

\subsubsection{Query View}
In \boxed{\textbf{A}} we see the query view, where a user can design a query consisting of player locations and game context to find similar game scenarios (\ref{req:query_context}). Drawing from sports play retrieval systems, we utilize a sketch-based interface \edit{\boxed{\textbf{A1}}}, whereby a user can mark player positions by clicking on a map. The user differentiates between T and CT sided players through sliders to indicate side. Additionally, a user can issue two kinds of spatial queries: (1) a full query, where every player is drawn or (2) a partial query, where only a subset of players are drawn. The full query is useful for situations where users want to identify how often a team utilizes certain global strategies, which is similar to the idea of scouting a team's ``formation'' in soccer. A full query is also useful for trying to find game states that may be known beforehand. The partial query is useful for situations when a user wants to analyze a partial region of the map, such as a bombsite. A user can mark if they want to perform a full or partial query through a slider. Once a player is placed on the map, the user can manipulate the placed player's position, which facilitates easy editing of game states. Such functionality also allows for rapid turnaround when a user tries to answer ``what if'' questions, such as what happens when a player assumes a different position on the map.

Below the spatial query view, a user can select from a variety of filters \edit{\boxed{\textbf{A2}}}. These filters are motivated by conversations our expert users, who oftentimes found themselves searching and filtering data often on specific teams, equipment values and number of grenades on each team. In order to facilitate any targeted analysis, the team filter is particularly important, as we gathered from our experts that most game analysis is spent with a particular opposition in mind. Furthermore, much of the analysis process is also spent understanding how teams operate under specific \textit{buy types}. Buy types are determined by the aggregate equipment values of a team. For example, if a team is saving money for a round and lacks the best equipment, they are described as being in an ``eco'' or ``semi buy'' situation. Alternatively, if a team starts the rounds with the best equipment, they are said to be in a ``full buy'' situation.


\subsubsection{Result View}
After a query is designed and executed from the query view, ggViz returns the set of results in \boxed{\textbf{B}}, the result view. Each element in the query's result set is displayed as a card which contains relevant information on the match and round in which the game state occurred, such as the match date, competition name, round score and buy types, and an icon indicating how the round ended (\ref{req:result_metadata}).

As the result set can be large for certain queries, users need a way to navigate through the result set to find interesting rounds. One approach to identify rounds with interesting moments is through analyzing \textit{win probability}. Win probability is a number which is an estimate for a side's chance of winning a given round. Effectively, we are estimating $\mathbb{P}(Y_r = 1 \mid G_{r,t})$, where $Y_r$ is the outcome of round $r$ (1 if the CT side wins, 0 otherwise), and $G_{r,t}$ is the state of the game in round $r$ at time $t$. In order to estimate this probability, we can learn some function $f(G_{r,t}) = \widehat{Y}_{r,t}$, where $\widehat{Y}_{r,t}$ is the estimated outcome of round $r$ at time $t$. To learn $f$, we use the methodology in \cite{XenopoulosWPACSGO} to train an XGBoost model which predicts a side's win probability at time $t$ given $G_{r,t}$.

In each result view card, we plot the win probability of the round in which the game state occurred \edit{\boxed{\textbf{B1}}}. We use a line chart which to visualize the win probability over the course of a round, where CT win probability is blue, and T win probability is orange. We indicate where in the round the game state occurred through a small circle at the top of the line chart. Finally, we also indicate important events in the round, such as when a bomb plant happened (black circle), which indicates frames that are ``pre-plant'' or ``post-plant'', since teams care about identifying ``post-plant'' situations. Essentially, the win probability chart presents a high-level overview of the round in which the game state occurred. Wallner~et~al. find that players are not only interested in low-level data, but also visualizations which represent high-level concepts~\cite{DBLP:conf/chi/WallnerWBK21}.

After the result view populates, a user can summarize the result set through a heatmap \edit{\boxed{\textbf{C}}}, which a user can select in the upper left corner of ggViz. To construct the heatmap, we calculate the density of one side's player locations in all game scenarios in the result set. Users can select whether they want to view the CT or the T position heatmap. By viewing the heatmap, users can identify common spots in which teams position themselves (\ref{req:heatmap}). Furthermore, by adjusting the filters in the query view, users can easily create different visualizations as content for other partner stakeholders, such as players receiving a pre-game presentation from a coach.


\subsubsection{Playback View}
When a user selects a game state of interest, the playback view in \boxed{\textbf{D}} populates. The playback view allows a user to view the round in which the game scenario occurred (\ref{req:playback}). Player positions are displayed on the map \edit{\boxed{\textbf{D1}}}. Below the map is a slider which one can use to animate the round frame by frame. This feature acts as a quick substitute for \edit{in-game replay} review, which typically takes place by playing back a demo file. Playing a demo file incurs a lengthy setup cost, as doing so typically requires running CSGO and loading a demo file. This greatly prohibits one from quickly switching between rounds in different demo files. However, if a user wants to view the demo file, we display a button to the right of the playback slider which will launch the CSGO game to watch the demo file.


By presenting the whole round for playback, a user can accomplish a variety of tasks. For example, a user can look at the game scenarios prior to selected one to develop a sense for how the selected game scenario came to be. On the other hand, a user can view the part of the round which occurs after the selected state to uncover what was the result of a particular scenario. Below the slider are tabs to view the round win probability, kills and grenades. In the win probability graph, we show added context for the round the user is reviewing. As major round events, like kills, are highly correlated with win probability~\cite{XenopoulosWPACSGO}, a user is quickly able to visit important round events. In the ``kill'' and ``grenades'' tab, a user can click to go to the frame where the kill occurred or grenade was thrown \edit{\boxed{\textbf{D2}}}. These specific player events add more context to the 2D view, and can help a user understand the circumstances of a given round.

\subsection{Implementation}
Early play retrieval systems, such as Chalkboarding, relied on non-web based interfaces. However, web-based applications are now ubiquitous, are able to support data-intensive workloads, such as play retrieval, and are easy to access for a wide array of users. As such, we prioritized ggViz to be accessible via a web-based client. ggViz is implemented with a client-server architecture. The client is a web application built with Angular, which renders the spatial query, result and round playback views. Using Flask, we develop an API which fulfills client requests, and in particular implements the \edit{aforementioned state} retrieval system. The API interacts with a SQL database that contains roughly 100 GB of data on 1,600 professional matches occurring from April 1st, 2020 to August 31st, 2020. This corresponds to over 10 million game states and over 100 million player locations. Game states are stored in a clustered SQL index, where the index is created on $(Map, Token)$ combinations. For the wide range of queries that our experts performed, each query produced a result set in just a few seconds. \edit{To search a single demo for a scenario of interest query would necessitate manually watching the demo, as no existing CSGO match review tool allows for scenario search. Since an average CSGO demo contains about 25 rounds, and if we assume an average round length of 1.5 minutes, finding all scenarios of interest would require almost 40 minutes.} We acquired the demo files from the popular Counter-Strike website HLTV.org. We parsed each demo file using the \texttt{awpy} CSGO demo parser at a frequency of one state per second~\cite{XenopoulosWPACSGO}. 

Both the client and server architectures are modular. The visual components, such as the playback view or query interface, can easily be extended to other systems. This is important for future systems, as most web-based visual analytics applications in esports will ostensibly require some 2D playback or map-based view. Additionally, our API can broadly support many CSGO specific applications, and can easily be extended by teams, organizations and players. For example, our API contains endpoints to retrieve event data, such as grenades, kills and damages, or player-tracking data. These endpoints can be used to retrieve data to power machine learning models or other visual analytics systems. Generally, these endpoints and data can also be used for 2D CSGO round playback.

%% file: 07_evaluation.tex
\section{Evaluation} \label{sec:evaluation}
To evaluate the usefulness of ggViz for esports analytics, we conduct expert interviews, where we showcased ggViz to staff from professional esports teams and recorded their feedback. Additionally, we showcase two use cases of ggViz, \edit{which we identified common analyses that CSGO experts perform in match analysis.}

\subsection{Expert Interviews}
\edit{Due to difficulty in recruiting experts with established esports presence, as well as extensive match review experience, we evaluated ggViz with the four esports experts mentioned in the pre-design interviews in section~\ref{sec:domain-req}, as well as an additional analyst, denoted A3.} A3 was part of the same esports organization as C1. Although we only used five experts, three of the five experts \edit{(A1, A2, M1)} were working for, at the time, top 10 ranked CSGO teams. 

\edit{We held ``think aloud'' interviews with the aforementioned experts. In our think-aloud protocol, experts were encouraged to vocalize their thoughts as they used ggViz.} During \edit{the interviews}, which \edit{each} lasted an hour and were held over Zoom, we first introduced ggViz and \edit{its} various components, and then recorded the expert's thoughts on ggViz as they used the system. \edit{ggViz was preloaded over 10 million game states parsed from professional CSGO demos.} The experts were free to use ggViz to accomplish whatever tasks they wanted. We note four broad categories in which we analyze expert use in retrieving game states: applying game context filters, sketching game states, viewing retrieval results and exploring tactics.

\subsubsection{Applying Game Context Filters}
Each expert's first actions with ggViz were to confirm tendencies for their own team. However, before drawing scenarios of interest, each user first inspected and applied the filters below the map query view. Of particular note were the team, round equipment and grenade count filters. Almost all queries that the users issued used these filters, which is consistent with the conclusions from the pre-design interviews. Before each query, each expert would vocalize what situation they wanted to find. In every instance, each expert referenced the ``buy type'' of a particular side, which is a rough hierarchy of equipment values. For example, an ``eco'' round is one where a team spends little to no money for a round in an effort to save money, and a ``full buy'' is one where a team acquires the best guns available. Our filters used the same nomenclature, which the experts appreciated. Overall, A2 remarked that the filtering was quite extensive and was satisfied that the filters corresponded to his team's general analysis workflows, saying ``the filters are exactly what we need''.

\subsubsection{Sketching Game States}
After selecting the filters, each expert would then draw the game state, which normally took under 10 seconds. All participants found the interface intuitive as the map view was familiar to anybody who has played the game. A1 mentioned that such an interface is ``super useful for coaches and for match preparation''. While ggViz allows a user to specify both CT and T players, most expert queries issued either partial queries, or full queries on a specific side. M1 and C1 said that he especially appreciated the ability to perform single-sided and partial queries, as a large part of their workflows include seeing how specific teams spatially distribute themselves on the CT side, regardless of how the T side is distributed. After later explaining the retrieval process to the experts, each user's drawing speed was much faster, since each expert understood that their game state drawings need not be exact, and they spent less time on putting players in exact locations. We also observed A1 and A2 try to understand the retrieval methodology further by purposely placing players near the borders of navigation mesh surfaces and seeing how the result set changed. 

\subsubsection{Viewing Retrieval Results}
The retrieved game states, along with the speed of the retrieval, were well received. M1 said that the similar plays looked ``great, they correspond to what I'm looking for''. Additionally, M1 mentioned that the query speed was well-suited to delivering fast turnaround for his workload. A2 remarked that the retrieval was ``fluid and fast, I don't have to wait forever to find what I want'' and that the retrieved game states made sense given all of his queries. C1 said ``I drew the exact setups I know my team uses, and I can see all of them within seconds. The returned states are excellent''. A3 remarked that the system is ``very accurate and returns the game states I expect, and at the point in the round I expect them to occur''. \
\edit{The experts varied in how they analyzed the result set and its corresponding rounds}. For example, C1 would quickly look at a few rounds (two to four), since his workflow was trying to understand how a team's CT sets up. C1 found the heatmap feature useful as well, since he could use these heatmaps to send to players. Conversely, A1 would spend a few minutes on a single game scenario by playing back the round's trajectories multiple times and jumping to different points in the round in order to understand the specific events of that round. Experts found that the win probability graph was useful in finding important points in a round as it clearly identified kills and deaths through large increases or decreases in win probability.

One concern of that the result view is that an excessive amount of results may be impossible to analyze. In practice, we observed that the experts followed two main paths of analysis. The first involved searching for a specific setup in a well-defined context that filters on a team and buy type. In these cases, it was rare for the result set to appear intractable to the experts. Furthermore, the experts would view a few rounds to confirm their hypothesis or to generate a new hypothesis, which would fall under the ``sense making'' and ``data validation'' categories of Kleinman~et~al.'s taxonomy~\cite{DBLP:journals/pacmhci/KleinmanPTBE21}. On the other hand, sometimes experts would relax their filters and queries to be less specific. In these cases, the heatmap view dominated analysis, rather than manual inspection of each game state. Thus, we did not find a strict need, nor did the experts express a desire, for specific ordering of the results.

\subsubsection{Exploring Tactics}
While team filters were present on almost all queries, A2 started to use the tool without a team filter towards the end of his interview. He suggested that although ggViz's direct use case is for game review, ggViz could also be used for a large scale analysis since the database of games was large and the retrieval speed was fast. A2 mentioned that most coaches and analysts in the CSGO community watch demos from their own team or their upcoming opposition, and thus lack the time to watch game replays in-depth from other top teams. With this in mind, ggViz could allow experts to learn more from other teams, beyond just the opposition, in a faster manner than going through individual demo files.

\subsubsection{Expert Feedback}
The feedback from the participants was very positive and each of them were excited to use the tool in depth and to suggest new features for future work. A2 said that the system solved his problem of finding specific game setups and contexts. C1 stated that he saw potential for ggViz to be ``more valuable than the other tools out there''. A2, C1 and M1 expressed interest that future versions of ggViz include more analytics, such as assigning values to player actions. One option is to assign values using the change in win probability, as is done in~\cite{XenopoulosWPACSGO}. The majority of the negative feedback of ggViz relied mostly on cosmetic details in the round playback view, such as what directions players were viewing or animations for grenades. One limitation of ggViz, which was mentioned by each expert, is how the 2D view is restrictive for dealing with maps that are vertically oriented. We elaborate on this issue in our discussion.

\subsection{Use Case: Analyzing Retakes}

\begin{figure}
    \centering
    \includegraphics[width=\textwidth]{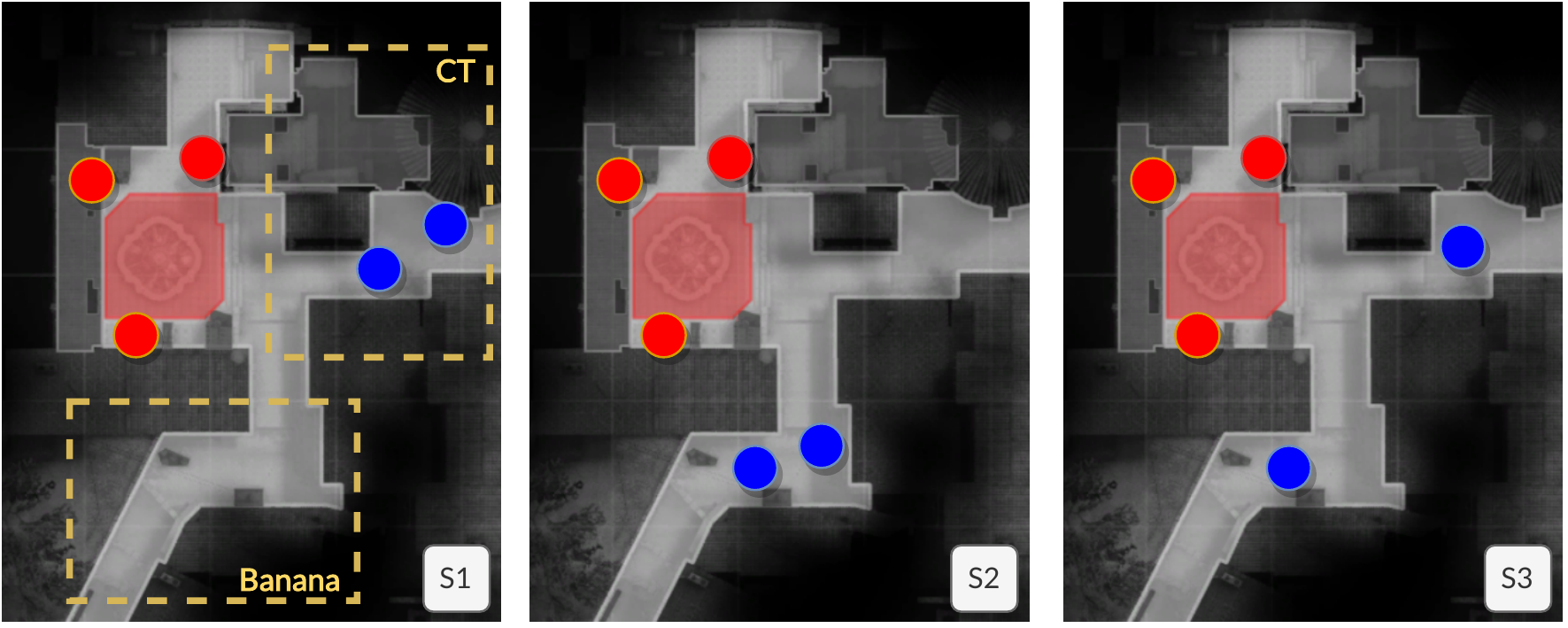}
    \Description[Retake analysis use case]{ggViz used to analyze 3 T vs 2 CT retake strategies on Inferno B bombsite.}
    \caption{S1, S2 and S3 detail three possible retake strategies for the Counter-Terrorists in a 2 CT versus 3 T scenario for the B bombsite on the \textit{Inferno} map. Retakes originating from ``Banana'' have a lower success rate than those from ``CT''. The banana and CT areas are labeled in S1.}
    \label{fig:retake-examples}
\end{figure}

C1 \edit{and A1 both} mentioned that \edit{they found ggViz useful for analyzing CT defensive setups and retake scenarios.} A \textit{retake} is a common maneuver in CSGO, whereby the CT sided team attempts to regain control of a bombsite where the T team has planted the bomb. Even though there is no definitive retake filter in ggViz, simply placing T players in common retake defense places and enforcing a bomb exploded or defused filter ensures that the returned states are mostly retakes.

Consider a case where a user is interested in retake strategies for a two CT versus three T situation for bombsite B on the map \textit{de\_inferno}. A two CT on three T situation is not uncommon, as usually two to three CTs defend a bombsite, so if a T side gains control of a bombsite, it is likely the other CT players were killed in the process. To isolate retake scenarios, we can filter for rounds that ended with the bomb exploding, the CT side being killed\edit{,} or the bomb being defused. In the former two scenarios, the T side wins the round. For the latter scenario, the CT side wins the round. In Figure~\ref{fig:retake-examples} we show three queried game situations\edit{, with two paths the CT players may take: ``banana'' or ``CT''. These situations vary on CT retake approaches.} Although a two CT versus three T retake is already a difficult situation for the CT side, \edit{using ggViz, it is clear S1 offers more advantage, as it attains a 12\% win rate compared to S2 (5\%) or S3 (0\%).}


\subsection{Use Case: Summarizing Team Setups} \label{sec:summarize-team-setups}
In the previous use case, we demonstrated the use of ggViz for understanding the events that occur \textit{after} a given game state. To do so, one must primarily use the result view and playback view. However, it is also important to summarize the result set itself. For example, a coach may want to create a visualization to share among players that describes how an opposing team defends a specific bombsite.

\begin{figure}
   \centering
    \includegraphics[width=\textwidth]{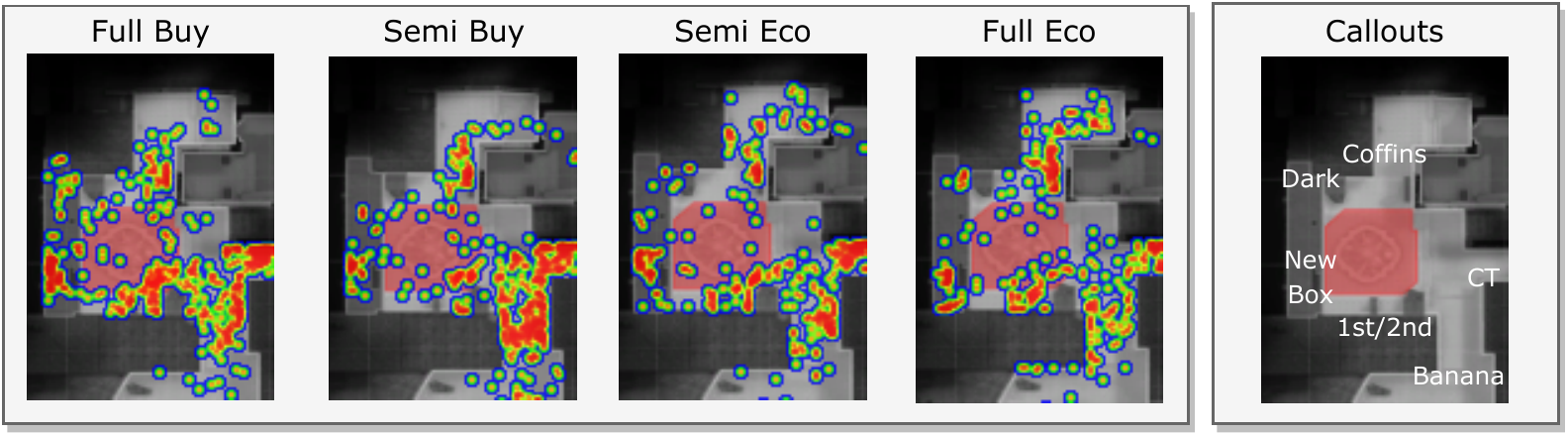}
    \Description[Summarizing team setups]{ggViz used to summarize team setups on Inferno B bombsite for different buy types.}
    \caption{We issue a partial query for two CT players on Inferno's B bombsite. We generate heatmaps for four different buy scenarios, and we provide the callouts for reference. We see that in full buy scenarios, the CT players are more likely to position themselves at 1st/2nd and Dark. In Semi Buys, players are more likely to position themselves closer to Banana. }
    \label{fig:heatmap}
\end{figure}

Consider the query in Figure~\ref{fig:heatmap}. Here, we issue a partial query that places two players on B bombsite in the map Inferno. Typically, two players defend the B bombsite on the CT side. We are interested in summarizing what positions these two players are likely to play. Furthermore, we are also interested how these positions differ under different buy conditions, such as a full buy or a semi-buy. This is a common task performed by coaches and analysts to find what they call a team's ``default''. In Figure~\ref{fig:heatmap}, we see the player position heatmaps for scenarios with two players on bombsite B under different buy scenarios. We provide the callouts for reference in the right side of the figure. We can see that there are clear positioning distinctions between the different buy types. For example, in full buys, players are more likely to position themselves at 1st/2nd and Dark. In semi buys, we see that players are likely to play closer to banana, and leaving Dark and 1st/2nd sparsely occupied. Interestingly, we see that in pistol rounds, players take positions deep in the bombsite, such as at Coffins or New Box.

%% file: 08_discussion.tex
\section{Discussion} \label{sec:discussion}

\edit{Our study provides insight as to how top-level esports coaches, managers and analysts review in-game performance.} We found that the workflows between the experts were quite similar and revolved around summarizing opponents' strategies across broad game scenarios, most commonly defined by the locations of players and the buy types of each team. Buy types were a fundamental dimension along which the experts started to group states and perform their analysis. Furthermore, there was a demonstrated need to find similar game states through player positions. Often, this need arose from a desire to group and analyze defensive setups of opponents, or to find exemplar game states to reference in analysis. Our work emphasizes the need for expanded esports data and analytical tools, particularly those directed towards the needs of expert esports stakeholders. \edit{Furthermore, ggViz presents a baseline for esports game state retrieval. In this section, we outline the limitations of our study, how it may be generalized to other esports, and directions for future work.}

\edit{ \subsection{Generalizability} }

\edit { 
Although we presented ggViz with a specific focus on CSGO, our backend and frontend should be easily extensible to other esports, provided two main components are available: player locations and a discretization of the 3D world in which the game is played. The latter can be created by a user arbitrarily. For example, editing a navigation mesh is something that is possible in some games; CSGO allows a player to generate and edit a navigation mesh through the game itself. In fact, the main limitation is in the former -- data acquisition. While we would have preferred to evaluate ggViz over a collection of FPS games, the data acquisition process is complicated by a lack of public game replay files and parsing software for many games. However, even if was only provided with player coordinates, one would easily be able to apply the state tokenization approach.

Since many FPS games have similar objectives and game design to CSGO, our approach and the results of our expert study may be useful for games like Valorant, Rainbow Six, Halo, Call of Duty or Overwatch -- each of which has an established esports community. However, significant game-specific consideration is required for designing non-spatial filters. For example, we derived the need for a ``buy type'' filter based on the experts tending to analyze teams through different buy types. This is a CSGO-specific filter, and may not apply to a game like Call of Duty or Halo. If we consider character-based games, like Valorant or Overwatch, some changes to the retrieval step may be necessary to account for positions not only of players but specific agents. Another consideration is the amount of variation in player positioning among games which feature different respawn characteristics or non-euclidean movement. For example, Halo features respawn-based competitive gameplay, where in CSGO a player has one life per round. Some games may also feature teleportation, such as Valorant. These properties may make player positioning more variable. The effects of these properties are unclear on the usefulness of ggViz, thus demonstrating the need for further evaluation of ggViz on other esports.
}

\edit{ \subsection{Limitations} }
One limitation of our navigation mesh-based approach is that when a player crosses a place border, the corresponding state token changes drastically, thus making the states dissimilar, while the underlying position is still similar. However, none of the experts commented or expressed concern on this issue. Another limitation of ggViz comes when dealing with maps that have varying height. For example, some maps in CSGO are especially vertically oriented. In situations like these, a single 2D view may be confusing. To address this, we can use multiple maps to display locations, where each map visualizes a certain height level. For example, in CSGO there exist ``upper'' and ``lower'' images for Nuke and Vertigo, which are two maps with large height differences. To display player positions on these upper and lower maps, there is typically a Z-axis threshold that if a player crosses, they are shown on switch from being visualized on one map to another.

\edit{ There are also limitations with regards to the evaluation of ggViz. Firstly, the sample size of four experts for the requirements gathering study, and five experts for the think-aloud evaluation, may be limiting. Although the evaluation utilizes a small sample of users, the sample is highly expert, as all experts work for highly-ranked teams with salaried rosters. Furthermore, since our experts span four esports organizations, it is unlikely that the observed expert workflows are unique to the idiosyncrasies of a particular organization. Another limitation of our evaluation is our observation-based approach instead of task-based approach. Given that there is no prior baseline system or set of tasks, and we have a small sample of users, a qualitative evaluation can lead to high-fidelity understanding of how CSGO analysts, coaches and managers may use ggViz, and can also help influence the design of future systems to be evaluated in a task-based manner.
}

\edit{ \subsection{Future Work} }

\edit{
ggViz as a system contains several avenues for future work. One of the main avenues of future work is extending our state retrieval framework to accommodate sequences of game states. A simple approach could involve treating periods of the game as ``sentences'', where each game state is a ``word'' that has an associated token. Similar trajectories may have similar subsequences, and one could use a metric such as the longest common subsequence to rank candidate states. Another area of future work is in the ranking of returned results in the game state retrieval process. Di~et.~al.~proposed a pairwise learning-to-rank approach for returned sports play results~\cite{di2018large}. They found that users favored results ranked through their approach. Although our users did not comment on the result set, nor felt overwhelmed by it, an improved ranking of game states may better facilitate game analysis.
}

%% file: 09_conclusion.tex
\section{Conclusion} \label{sec:conclusion}
This paper presents ggViz, a visual analytics system designed to query similar game states from a large corpus of spatiotemporal CSGO data. Informed through interviews with top CSGO coaches and analysts, we design ggViz to specifically aid game review and tactic discovery. ggViz allows users to search for similar game situations through a sketch-based query interface. Underlying ggViz is a performant game state retrieval framework that creates tokens which represent player locations. ggViz supports both full and partial spatial queries. We evaluate ggViz through expert interviews, as well as through use cases motivated by expert use. Our evaluation suggests that ggViz can greatly enhance the game review and tactic discovery process for esports analysts, coaches, and managers by decreasing the time it takes to find game states of interest. Furthermore, our approach is easily extensible to other esports.